\documentclass[reprint,superscriptaddress,
 amsmath,amssymb,
 aps,
]{revtex4-1}

\usepackage{graphicx}
\usepackage{dcolumn}
\usepackage{bm}
\usepackage{dcolumn}

\begin{document}

\preprint{APS/123-QED}

\title{Precursor of a magnetic-field-induced liquid-liquid transition of oxygen}
\author{T. Nomura}
\email{t.nomura@issp.u-tokyo.ac.jp}
\affiliation{Institute for Solid State Physics, University of Tokyo, Kashiwa, Chiba 277-8581, Japan}
\affiliation{Hochfeld-Magnetlabor Dresden (HLD-EMFL), Helmholtz-Zentrum Dresden-Rossendorf, 01328 Dresden, Germany}
\author{Y. H. Matsuda}
\email{ymatsuda@issp.u-tokyo.ac.jp} 
\affiliation{Institute for Solid State Physics, University of Tokyo, Kashiwa, Chiba 277-8581, Japan}
\author{S. Zherlitsyn}
\affiliation{Hochfeld-Magnetlabor Dresden (HLD-EMFL), Helmholtz-Zentrum Dresden-Rossendorf, 01328 Dresden, Germany}
\author{J. Wosnitza}
\affiliation{Hochfeld-Magnetlabor Dresden (HLD-EMFL), Helmholtz-Zentrum Dresden-Rossendorf, 01328 Dresden, Germany}
\affiliation{Institut f\"ur Festk\"orper- und Materialphysik, Technische Universit\"at Dresden, 01062 Dresden, Germany}
\author{T. C. Kobayashi}
\affiliation{Department of Physics, Okayama University, Okayama 700-8530, Japan}

\date{\today}

\begin{abstract}
The acoustic properties of liquid oxygen have been studied up to 90 T by means of the ultrasound pulse-echo technique.
A monotonic decrease of the sound velocity and an asymptotic increase of the acoustic attenuation are observed by applying magnetic fields.
An unusually large acoustic attenuation, that becomes 20 times as large as the zero-field value, cannot be explained by the classical theory.
These results indicate strong fluctuations of antiferromagnetically coupled local structures.
We point out that the observed fluctuations are a precursor of a liquid-liquid transition, from a low-susceptibility to a high-susceptibility liquid, which is characterized by a local-structure rearrangement.
\end{abstract}
\maketitle

\section{introduction}
Classically, the liquid phase is treated as a homogeneous state of matter where molecules are disordered translationally and orientationally.
In this theoretical framework, the liquid and gas phases have the same symmetry and the density difference is taken as the order parameter to distinguish them.
In the last two decades, discoveries of liquid-liquid transitions (LLTs) have opened a new perspective on the classical understanding of liquids \cite{Poole1997,Tanaka2000,McMillan2007,Gallo2016}.
Two distinguishable states of a liquid indicate that the symmetry of the liquid breaks locally and dynamically.
LLTs have been proposed for many elemental and molecular liquids at high pressure, and are recognized as ubiquitous phenomena \cite{Katayama2000,Cadien2013,Boates2009,Sastry2003,Lorenzen2010,Jara2009,Tanaka2004,Mishima1998}.

Up to now, LLTs have been mostly discussed as a function of pressure, $P$, and temperature, $T$ \cite{Anisimov2018}.
For the case of H$_2$O, a low-density liquid (LDL) and high-density liquid (HDL) are proposed \cite{Gallo2016}.
The frustration between hydrogen-bond (directional) and van der Waals interaction (isotropic) results in two competing states, a tetrahedrally coordinated LDL and close-packed HDL \cite{Tanaka2000,McMillan2007,Gallo2016}.
Pressure tunes the delicate balance between them and induces a crossover (possibly a phase transition in the supercooled regime).
Here, LDL and HDL are distinguished by density and bond-order parameters; the latter is defined as the local fraction of locally favored structures \cite{Tanaka2000}.

When the molecule is magnetic, it is natural to introduce the magnetic field, $H$, as a thermodynamic variable, implying that $H$ can also drive a LLT.
In that case, the magnetic susceptibility $\chi$ is taken as the primary order parameter, and two states of the liquid would be distinguished as low-susceptibility liquid (L$\chi$L) and high-susceptibility liquid (H$\chi$L).
A promising candidate for the magnetic-field-induced LLT is liquid oxygen.
The O$_2$ molecule has a spin $S=1$, and liquid oxygen (90.2--54.4 K) behaves as a paramagnetic liquid with antiferromagnetic (AFM) correlations \cite{Lewis1924,Kratky1975}.
The Curie-Weiss temperature is approximately $-50$ K \cite{Uyeda1988,Brodyanskii1989}.
The AFM exchange interaction leads to the dynamical dimerization of O$_2$ molecules with a singlet spin state, which is  suggested by magnetic-susceptibility \cite{Kratky1975,Lewis1924}, optical absorption-spectra \cite{Tsai1969,Uyeda1988,Bhandari1973,Landau1961}, and neutron-scattering data \cite{Fernandez2008,Chahid1993}, as well as by molecular dynamics (MD) simulation \cite{Oda2002,Oda2004}.

In fact, an external magnetic field tunes the geometrical alignment of the O$_2$-O$_2$ dimer \cite{Hemert1983,Bussery1993,Bartolomei2008,Obata2013}.
At zero field, the H geometry (rectangular parallel) is the most stable alignment which maximizes the AFM exchange interaction with maximized overlap integral.
When the magnetic moment is fully polarized by external fields, the O$_2$-O$_2$ dimer prefers X (crossed) or S (canted) geometries to minimize the overlap integral.
Such a field-induced molecular rearrangement has been observed in the solid oxygen $\alpha$-$\theta$ phase transition at around 100 T \cite{Nomura2014,Nomura2015,Nomura2017PD}.
The same mechanism would be relevant also for a LLT of oxygen where the favored molecular alignment (H, X, S, distinguished by the bond-order parameter) changes by magnetic field.

In this paper, we present the ultrasonic properties of liquid oxygen up to 90 T.
The experimental results, which indicate strong fluctuation of the local structures, are discussed in the context of the proposed magnetic-field-induced LLT.

\section{Experiment}
\begin{figure}[ptb]
\centering
\includegraphics[width=8.6cm]{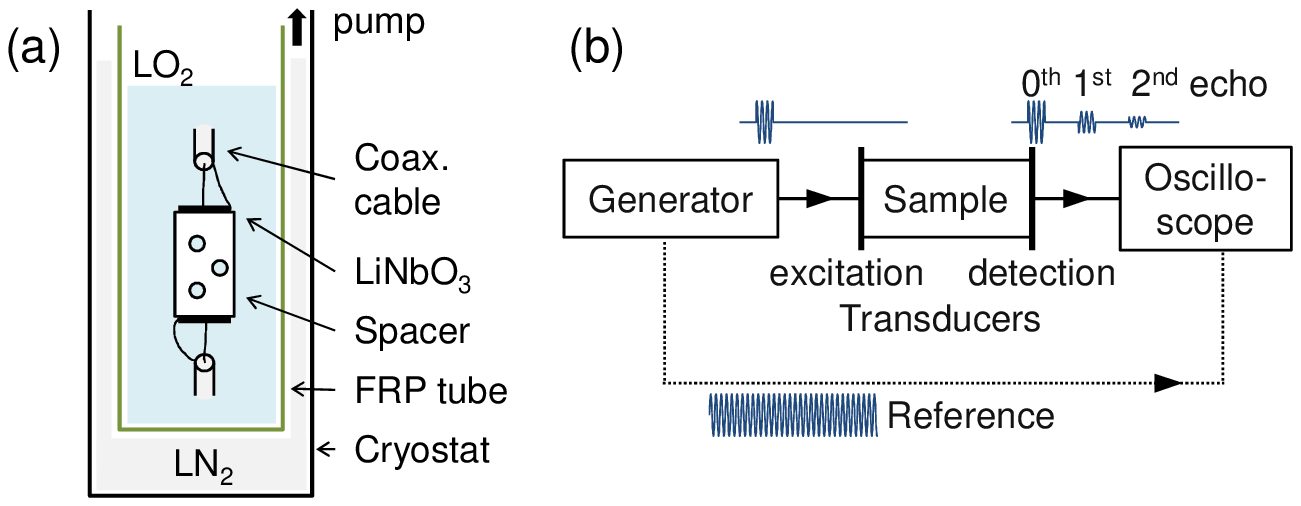}
\caption{\label{fig:exp}
(a) Experimental configuration of the ultrasound measurement for liquid oxygen.
(b) Block diagram of the ultrasound pulse-echo technique.
}
\end{figure}

An ultrasound pulse-echo technique with digital homodyne detection was used to study the sound velocity, $v$, and acoustic-attenuation coefficient, $\alpha$.
A scheme of the experiment is shown in Fig. \ref{fig:exp}.
Two LiNbO$_3$ transducers (Y-36$^\mathrm{o}$ cut, fundamental resonance at 20--40 MHz) were placed parallel by using a plastic spacer (typically, sample length $L=5$ mm).
The sample space was filled by pure oxygen gas (99.999 \%).
On cooling, liquid oxygen with vapor pressure was condensed at the bottom of a fiber-reinforced plastic (FRP) tube.
Condensed oxygen was automatically filled in the plastic spacer through the holes.
Radio frequency (RF, 20--210 MHz) acoustic-wave pulses were excited and detected by the transducers.
The echo signal and reference RF were recorded by a digital oscilloscope (LeCroy, HDO 8108A).
The phase and amplitude of the 0th echo were detected by digital homodyne and converted to the relative change of $v$ and $\alpha$ as,
\begin{equation}
\Delta v/v=\Delta \Phi/\Phi=\Delta \Phi/(2\pi f \tau_0),
\label{eq:vel_phase}
\end{equation}
\begin{equation}
\Delta \alpha=-\mathrm{ln}(A/A_0)/L.
\label{eq:att_amp}
\end{equation}
Here, $\Phi$ is the phase of the detected echo, $f$ is the ultrasound frequency, $\tau_0$ is the travel time of the sound for $L$, and $A/A_0$ is the amplitude normalized by zero-field value.
The sound velocity at zero field is calibrated by the value from literature \cite{Itterbeek1962,Dael1966,Clouter1973}.
$v$ does not depend on $f$ below 3.6 GHz \cite{Clouter1973}.

Magnetic fields below 60 T were generated by a pulsed magnet with pulse duration of 150 ms.
The 90 T pulse was generated by a dual-pulse magnet (with the temporal field profile shown in the inset of Fig. \ref{fig:attenuation}) \cite{Zherlitsyn2013}.
The magnetic field was measured by a pickup coil placed near the plastic spacer.
The magnetic field was parallel to the ultrasound propagation direction.
The temperature immediately before the pulse was measured by a RuO$_2$ resistance thermometer.

\section{Results}
\begin{figure}[ptb]
\centering
\includegraphics[width=7.4cm]{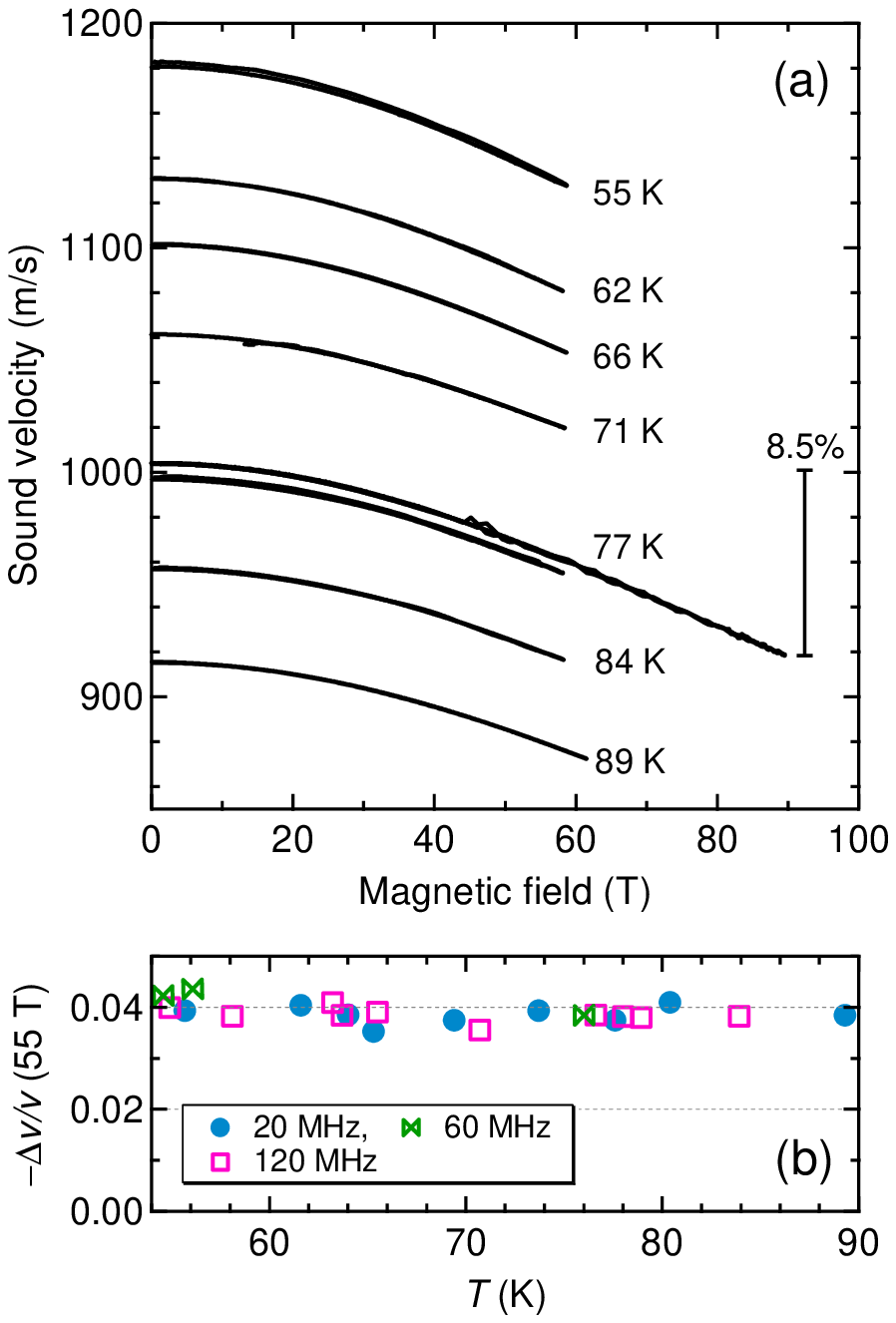}
\caption{\label{fig:velocity}
(a) Sound velocity of liquid oxygen as a function of magnetic field.
Temperature is denoted for each curve.
(b) Relative change of the sound velocity at 55 T as a function of temperature for each ultrasound frequency.
}
\end{figure}

Figure \ref{fig:velocity}(a) shows the sound velocity as a function of magnetic field for various temperatures.
In the whole temperature range, a monotonic decrease of $v$ is observed without any hysteresis.
The reversible results indicate that any	 temperature change (e.g., due to a magnetocaloric effect) is negligible in this experimental setup (see also Ref. \cite{Nomura2017MCE}).
At 90 T, $v$ decreases by 8.5 \% without any sign of saturation.
Figure \ref{fig:velocity}(b) shows the temperature and frequency dependence of the relative change of the sound velocity $\Delta v/v$ at 55 T.
Within our measurement accuracy, $\Delta v/v$ does not depend on $T$ and $f$.

\begin{figure}[ptb]
\centering
\includegraphics[width=8.0cm]{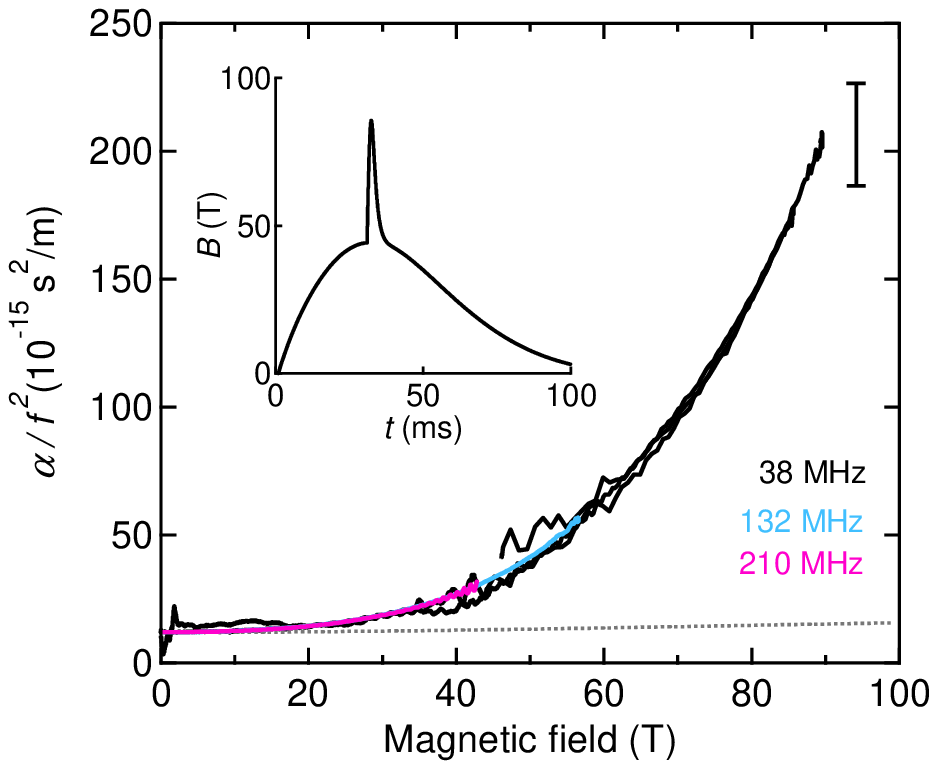}
\caption{\label{fig:attenuation}
Normalized acoustic attenuation coefficient $\alpha/f^2$ of liquid oxygen at 77 K as a function of magnetic field.
The error bar at 90 T is shown.
The dotted line shows the classically expected behavior based on Eq. (\ref{eq:expected_att}).
The inset shows the time-dependent magnetic-field profile.
}
\end{figure}

Figure \ref{fig:attenuation} shows the acoustic attenuation coefficient at 77 K up to 90 T.
Here, we plot $\alpha$ normalized by the square of the ultrasound frequency, $\alpha/f^2$, which gives one universal curve for the different ultrasound frequencies.
In this paper, we use in the following the unit $10^{-15}$s$^2$/m for the normalized attenuation.
Up to 90 T, $\alpha/f^2$ shows  a monotonous increase and reaches a value above 200, which is 20 times as large as the zero-field value.
Empirically, $\alpha/f^2$ is known to be approximately 10 for the simple non-associated liquids (Table \ref{tab:table1}).
This evidences that the observed value of 200 is extraordinarily large.

\begin{table}
\caption{\label{tab:table1}Normalized acoustic attenuation coefficients $\alpha/f^2$ for simple liquids.}
\begin{ruledtabular}
\begin{tabular}{lD{.}{.}{5}D{.}{.}{5}r}
Liquid & T\ (\mathrm{K}) & \alpha/f^2\ (10^{-15}\ \mathrm{s}^2/\mathrm{m}) & Ref. \\
\hline
O$_2$ & 77 & 11 & \cite{Victor1970} \footnote{zero-field value}\\
N$_2$ & 73.9 & 10.6 & \cite{Galt1948} \\
H$_2$ & 17 & 5.6 & \cite{Galt1948} \\
Ar & 85.2 & 10.1 & \cite{Galt1948} \\
H$_2$O & 353 & 10.6 & \cite{Hall1948}\\
H$_2$O & 278 & 61.4 & \cite{Hall1948}\\
\end{tabular}
\end{ruledtabular}
\end{table}

\section{Discussion}
The classical acoustic attenuation due to viscosity and heat conduction is described by \cite{Victor1970,Bhatia}
\begin{equation}
\frac{\alpha}{f^2}=\frac{2\pi^2}{\rho v^3}\left(\frac{4}{3}\eta_s+\eta_v+\frac{(\gamma-1)\kappa}{c_p}\right).
\label{eq:norm_att}
\end{equation}
Here, $f$ is the ultrasound frequency, $\rho$ is the density, $\eta_s$ is the shear viscosity, $\eta_v$ is the bulk viscosity, $\gamma$ is the heat capacity ratio, $\kappa$ is the thermal conductivity, $c_p$ is the specific heat at constant pressure.
Typical values are summarized in Table \ref{tab:table2}.
The acoustic attenuation dominantly originates from the viscosity ($\eta_s$ and $\eta_v$), and the contribution from the heat conduction is around 10~\% \cite{Victor1970}.

\begin{table*}
\caption{\label{tab:table2} Typical parameters of liquid oxygen at zero field \cite{Victor1970} and their contributions to the acoustic attenuation.}
\begin{ruledtabular}
\begin{tabular}{ccccccc}
$T$ & $\eta_s$ & $\eta_v$ & $\gamma$ & $\kappa$ & $c_p$  & 4/3$\eta_s$ : $\eta_v$ : ($\gamma-1$)$\kappa$/$c_p$ \\
(K) & (10$^{-3}$ Pa$\cdot$s) & (10$^{-3}$ Pa$\cdot$s) &  & (J/m$\cdot$s$\cdot$K) & (J/kg$\cdot$K) & \\
\hline
80.1 $\pm$ 0.2 & 0.245 & 0.212 & 1.687 & 0.163 & 1680 & 0.54 : 0.35 : 0.11\\
75.4 $\pm$ 0.2 & 0.282 & 0.279 & 1.678 & 0.167 & 1670 & 0.52 : 0.39 : 0.09\\
\end{tabular}
\end{ruledtabular}
\end{table*}

Assuming that the value inside the bracket in Eq. (\ref{eq:norm_att}) is field independent, the relative change of $\alpha$ is given by
\begin{equation}
\Delta\alpha/\alpha_0=-\Delta\rho/\rho-3\Delta v/v.
\label{eq:expected_att}
\end{equation}
The density decreases by 0.02~\% at 8~T \cite{Uyeda1987}, and a quadratic extrapolation gives $-2.8$~\% at 90~T.
Using the measured $\Delta v/v$ [Fig.~\ref{fig:velocity} (a)], the field dependence of the sound attenuation can be calculated and is shown as the dotted line in Fig.~\ref{fig:attenuation}.
$\Delta \alpha/f^2$ is at most 3, which is much less than the observed value.
Let us comment on our assumption of the field independence of the term in the bracket of Eq. (\ref{eq:norm_att}).
The dominant contribution of this term comes from $\eta_s$ and $\eta_v$.
$\eta_s$ of liquid oxygen decreases by 5 \% at 50 T \cite{Comment_visc}, which even reduces the change of $\alpha$.
To explain the observed attenuation based on Eq. (\ref{eq:norm_att}), $\eta_v$ would have to become 50 times larger than the zero-field value, which is not realistic.
Thus, the classical attenuation mechanisms due to viscosity and heat conduction cannot account for the observed acoustic attenuation.

As possible reasons for the excessive acoustic attenuation, (i) relaxation and (ii) a phase transition (including critical phenomena) are proposed \cite{Bhatia}.
The effect of relaxation is observed when the ultrasound frequency is close to the relaxation dynamics of the molecules.
As an estimate for that the so-called Lucas relaxation frequency is used \cite{Fleury1969},
\begin{equation}
f_\mathrm{L}=\frac{\rho v^2}{4/3\eta_s+\eta_v}\sim 2\ \mathrm{THz}.
\label{eq:Lucas}
\end{equation}
The parameters are taken from Table \ref{tab:table2} \cite{Victor1970}.
$f_\mathrm{L}$ is consistent with the relaxation time obtained from inelastic neutron scattering \cite{Fernandez2008} and is orders of magnitude larger than the ultrasound frequencies used in this study.
Moreover, $\Delta v/v$ and $\Delta \alpha/f^2$ do not depend on the ultrasound frequencies (Figs. 2 and 3), indicating the dispersion is negligible in this frequency range.
Therefore, we can exclude the effect of relaxation.

As a possible explanation for the excessive acoustic attenuation we, therefore, propose a nearby phase transition (with corresponding critical phenomena), from L$\chi$L to H$\chi$L.
The locally favored structure changes from H- to X- or S-type geometry by applying a magnetic field \cite{Hemert1983,Bussery1993,Bartolomei2008,Obata2013}, which results in a phase transition for solid oxygen at around 100 T \cite{Nomura2014,Nomura2015,Nomura2017PD}.
Near the phase boundary (or critical point), the local molecular arrangement is more degenerated and fluctuates.
The fluctuation of the local structure and density results in the strong acoustic attenuation and the decrease of sound velocity.
This is generally observed near a gas-liquid critical point \cite{Garland1970,Thoen1974,Garland1970_2}.
For the case of liquid H$_2$O, the increase of $\Delta \alpha/f^2$ by lowering the temperature (see Table \ref{tab:table1}) is explained in the context of two competing local molecular arrangements \cite{Hall1948}.
The observed increase of $\Delta \alpha/f^2$ would be also attributed to field-induced local-structure fluctuations, namely, a precursor of the field-induced LLT.

When the fluctuations are suppressed at extremely high fields ($\sim$300 T), $\alpha$ should show reasonable values explainable in the framework of the classical theory ($\sim$10).
Therefore, a maximum of $\alpha$ is expected at higher magnetic fields, where the LLT or a crossover takes place.
Thermodynamic analysis suggests that a second-order phase boundary is allowed for the vector order parameter $\Delta \bf M$, but not for the scalar order parameter $\Delta \rho$ \cite{Anisimov2018}.
In this sense, the magnetic-field-induced LLT is qualitatively different from the pressure-induced LLT.
Experiments at higher fields are needed to prove the existence of the magnetic-field-induced LLT.

\section{Conclusion}
Our high-field ultrasound measurements on liquid oxygen revealed that (i) $v$ decreases and (ii) $\alpha$ increases monotonously up to 90 T.
Especially, the observed acoustic attenuation cannot be explained by classical attenuation mechanisms.
These features reflect local-structure fluctuations, which are considered to be a precursor of a field-induced LLT.
The analogy with the solid oxygen $\alpha$-$\theta$ phase transition suggests that the LLT is driven by a field-induced molecular rearrangement.
Higher-field experiments are needed to confirm the proposed LLT.

Our finding, the precursor of a magnetic-field-induced LLT, suggests that LLTs are more general phenomena than generally assumed.
Using the three independent tuning parameters ($T$, $P$, and $H$) one may speculate that more than two liquid phases can be realized even for elemental liquids.
In this context, searching for a pressure-induced LLT for oxygen, which has not been proposed, is a promising venue for a deeper understanding of this liquid.

\section*{acknowledgments}
We thank O. Yamamuro, T. Oda, and M. Obata for fruitful discussions.
We acknowledge the support of the HLD at HZDR, member of the European Magnetic Field Laboratory (EMFL).
T. N. was supported by a Grant-in-Aid for JSPS Fellows.
This work was partly supported by JSPS KAKENHI, Grant-in-Aid for Scientific Research (B) (16H04009).


\begin{thebibliography}{99}
\bibitem{Poole1997}
P. H. Poole, T. Grande, C. A. Angell, and P. F. McMillan, Science {\bf275}, 322 (1997).
\bibitem{Tanaka2000}
H. Tanaka, Phys. Rev. E {\bf62}, 6968 (2000).
\bibitem{McMillan2007}
P. F. McMillan, M. Wilson, M. C. Wilding, D. Daisenberger, M. Mezouar, and G. Neville Greaves, J. Phys. Condens. Matter {\bf19}, 415101 (2007).
\bibitem{Gallo2016}
P. Gallo, K. Amann-Winkel, C. A. Angell, M. A. Anisimov, F. Caupin, C. Chakravarty, E. Lascaris, T. Loerting, A. Z. Panagiotopoulos, J. Russo, J. A. Sellberg, H. E. Stanley, H. Tanaka, C. Vega, L. Xu, and L. G. M. Pettersson, Chem. Rev. {\bf116}, 7463 (2016).
\bibitem{Katayama2000}
Y. Katayama, T. Mizutani, W. Utsumi, O. Shimomura, M. Yamakata, and K. I. Funakoshi, Nature (London) {\bf403}, 170 (2000).
\bibitem{Cadien2013}
A. Cadien, Q. Y. Hu, Y. Meng, Y. Q. Cheng, M. W. Chen, J. F. Shu, H. K. Mao, and H. W. Sheng, Phys. Rev. Lett. {\bf110}, 125503 (2013).
\bibitem{Boates2009}
B. Boates and S. A. Bonev, Phys. Rev. Lett. {\bf102}, 015701 (2009).
\bibitem{Sastry2003}
S. Sastry and C. A. Angell, Nat. Mater. {\bf2}, 739 (2003).
\bibitem{Lorenzen2010}
W. Lorenzen, B. Holst, and R. Redmer, Phys. Rev. B {\bf82}, 195107 (2010).
\bibitem{Jara2009}
D. A. C. Jara, M. F. Michelon, A. Antonelli, and M. de Koning, J. Chem. Phys. {\bf130}, 221101 (2009).
\bibitem{Tanaka2004}
H. Tanaka, R. Kurita, and H. Mataki, Phys. Rev. Lett. {\bf92}, 4 (2004).
\bibitem{Mishima1998}
O. Mishima and H. E. Stanley, Nature (London) {\bf396}, 329 (1998).
\bibitem{Anisimov2018}
M. A. Anisimov, M. Du\ifmmode \check{s}\else \v{s}\fi{}ka, F. Caupin, L. E. Amrhein, A. Rosenbaum, and R. J. Sadus, Phys. Rev. X {\bf8}, 011004 (2018).
\bibitem{Lewis1924}
G. N. Lewis, J. Am. Chem. Soc. {\bf46}, 2027 (1924).
\bibitem{Kratky1975}
K. W. Kratky, Acta Math. Acad. Sci. Hung. {\bf39}, 15 (1975).

\bibitem{Brodyanskii1989}
A. P. Brodyanskii, Yu. A. Freiman, and A. Jezowski, J. Phys.: Condens. Matter {\bf1}, 999 (1989).
\bibitem{Uyeda1988}
C. Uyeda, A. Yamagishi, and M. Date, J. Phys. Soc. Jpn. {\bf57}, 3954 (1988).
\bibitem{Tsai1969}
S. C. Tsai and G. W. Robinson, J. Chem. Phys. {\bf51}, 3559 (1969).

\bibitem{Bhandari1973}
R. Bhandari and L. M. Falicov, J. Phys. C: Solid State Phys. \textbf{6}, 479 (1973).
\bibitem{Landau1961}
A. Landau, E. J. Allin, and H. L. Welsh, Spectrochimica Acta \textbf{18}, 1-19 (1961).

\bibitem{Fernandez2008}
F. Fernandez-Alonso, F. J. Bermejo, I. Bustinduy, M. A. Adams, and J. W. Taylor, Phys. Rev. B {\bf78}, 104303 (2008).
\bibitem{Chahid1993} 
A. Chahid, F. J. Bermejo, E. Enciso, M. Garcia-Hernandez, and J. L. Martinez, J. Phys.: Condens. Matter {\bf5}, 423 (1993).


\bibitem{Oda2002}
T. Oda and A. Pasquarello, Phys. Rev. Lett. {\bf89}, 197204 (2002).
\bibitem{Oda2004}
T. Oda and A. Pasquarello, Phys. Rev. B {\bf70}, 134402 (2004).
\bibitem{Hemert1983}
M. C. van Hemert, P. E. S. Wormer, and A. van der Avoird, Phys. Rev. Lett. {\bf51}, 1167 (1983).
\bibitem{Bussery1993}
B. Bussery and P. E. S.Wormer, J. Chem. Phys. {\bf99}, 1230 (1993).
\bibitem{Bartolomei2008}
M. Bartolomei, M. I. Hernandez, J. Campos-Martinez, E. Carmona-Novillo, and R. Hernandez-Lamoneda, Phys. Chem. Chem. Phys. {\bf10}, 5374 (2008).
\bibitem{Obata2013}
M. Obata, M. Nakamura, I. Hamada, and T. Oda, J. Phys. Soc. Jpn. {\bf82}, 093701 (2013).
\bibitem{Nomura2014}
T. Nomura, Y. H. Matsuda, S. Takeyama, A. Matsuo, K. Kindo, J. L. Her, and T. C. Kobayashi, Phys. Rev. Lett. {\bf112}, 247201 (2014).
\bibitem{Nomura2015}
T. Nomura, Y. H. Matsuda, S. Takeyama, A. Matsuo, K. Kindo, and T. C. Kobayashi, Phys. Rev. B {\bf92}, 064109 (2015).
\bibitem{Nomura2017PD}
T. Nomura, Y. H. Matsuda, and T. C. Kobayashi, Phys. Rev. B {\bf96}, 054439 (2017).

\bibitem{Itterbeek1962}
A. Van Itterbeek and W. Van Dael, Physica {\bf28}, 861 (1962).
\bibitem{Dael1966}
W. Van Dael, A. Van Itterbeek, A. Cops, and J. Thoen, Physica {\bf32}, 611 (1966).
\bibitem{Clouter1973}
M. J. Clouter and H. Kiefte, J. Chem. Phys. {\bf59}, 2537 (1973).
\bibitem{Zherlitsyn2013}
S. Zherlitsyn, B. Wustmann, T. Herrmannsd\"orfer, and J. Wosnitza, J. Low Temp. Phys. {\bf170}, 447 (2013).
\bibitem{Nomura2017MCE}
T. Nomura, Y. Kohama, Y. H. Matsuda, K. Kindo, and T. C. Kobayashi, Phys. Rev. B \textbf{95}, 104420 (2017).

\bibitem{Victor1970}
A. E. Victor and R. T. Beyer, J. Chem. Phys. {\bf52}, 1573 (1970).
\bibitem{Galt1948}
J. K. Galt, J. Chem. Phys. {\bf16}, 505 (1948).
\bibitem{Hall1948}
L. Hall, Phys. Rev. {\bf73}, 775 (1948).
\bibitem{Bhatia}
B. Bhatia, {\it Ultrasonic Absorption: An Introduction to the Theory of Sound Absorption and Dispersion in Gases, Liquids, and Solids.} (Oxford University Press, New York, 1986).
\bibitem{Uyeda1987}
C. Uyeda, A. Yamagishi, and M. Date, J. Phys. Soc. Jpn. {\bf56}, 3444 (1987).
\bibitem{Comment_visc}
T. Nomura, S. Zherlitsyn, Y. Kohama, and J. Wosnitza, Rev. Sci. Instrum. {\bf90}, 065101 (2019).
\bibitem{Fleury1969} 
P. A. Fleury and J. P. Boon, Phys. Rev. {\bf186}, 244 (1969).

\bibitem{Garland1970} 
C. W. Garland, D. Eden, and L. Mistura, Phys. Rev. Lett. {\bf25}, 1161 (1970).
\bibitem{Thoen1974} 
J. Thoen and C. W. Garland, Phys. Rev. A {\bf10}, 1311 (1974).
\bibitem{Garland1970_2}  
C. W. Garland, in Physical Acoustics edited by W. P. Mason and R. N. Thurston (Academic, New York, 1970), Vol. 7, Chap. 2.


\end{thebibliography}
\end{document}